\documentclass[12pt]{article}
\usepackage{amsmath}
\usepackage{amssymb}
\usepackage{cancel}
\usepackage{graphicx}
\usepackage{wrapfig}
\renewcommand{\thefootnote}{\arabic{footnote}}

\numberwithin{equation}{section}

\newcommand{\nue}{\ensuremath{\nu_e} }
\newcommand{\numu}{\ensuremath{\nu_\mu} }

\newcommand{\antinue}{\ensuremath{\bar{\nu}_e} }
\newcommand{\antinumu}{\ensuremath{\bar{\nu}_\mu} }
\newcommand{\antinutau}{\ensuremath{\bar{\nu}_\tau} }

\begin{document}
\baselineskip=19pt

\begin{titlepage}

\begin{center}
\vspace*{17mm}

{\large\bf%
The Glashow resonance at IceCube: signatures, event rates
and $pp$ vs.~$p\gamma$ interactions
}

\vspace*{10mm}
Atri~Bhattacharya$^{*}$\footnote{~atri@hri.res.in},   
~Raj~Gandhi$^{*}$\footnote{~nubarnu@gmail.com},
~Werner~Rodejohann$^{\dag}$\footnote{~werner.rodejohann@mpi-hd.mpg.de},
~Atsushi~Watanabe$^{\dag, \ddag}$\footnote{~watanabe@muse.sc.niigata-u.ac.jp}
\vspace*{10mm}

$^*${\small {\it 
Harish-Chandra Research Institute, Chhatnag Road, Jhunsi, 
Allahabad 211 019, India
}} \\

$^\dag${\small {\it Max-Planck-Institut f\"ur Kernphysik,
  Postfach 103980, 69029 Heidelberg, Germany
}}\\

$^\ddag${\small {\it Department of Physics, 
Niigata University, Niigata 950-2181, Japan}}\\

\vspace*{3mm}

{\small (August, 2011)}
\end{center}

\vspace*{7mm}

\begin{abstract}\noindent%
We revisit the signatures of the Glashow resonance process $\bar{\nu}_e e \to W $ in the
high-energy astrophysical neutrino observatory IceCube. 
We note that in addition to the standard hadronic and electromagnetic showers produced 
by an incoming neutrino at the resonance energy of $E_\nu \approx 6.3$ PeV, 
there are two clear signals of the process: the ``pure muon" from 
$\bar{\nu}_e e \to \bar{\nu}_\mu \mu$ and the ``contained lollipop"
from $\bar{\nu}_e e \to \bar{\nu}_\tau \tau$. 
The event rate and the signal-to-background ratio (the ratio of the resonant 
to concurrent non-resonant processes) are calculated for each type of interaction, 
based on current flux limits on the diffuse neutrino flux.
Because of the low background in the neighborhood of the resonance, 
the observation of only one pure muon or contained lollipop event essentially 
signals discovery of the resonance, even if the expected event numbers are small.
We also evaluate the total event rates of the Glashow resonance from 
the extra-galactic diffuse neutrino flux and emphasize its utility as a discovery tool 
to enable first observations of such a flux. 
We find that one can expect 3.6 (0.65) events per year for a pure $pp$ ($p\gamma$) 
source, along with an added contribution of 0.51 (0.21) from non-resonant events. 
We also give results as a function of the ratio of $pp$ vs $p\gamma$ sources. 
\end{abstract} 

\end{titlepage}

\newpage
\renewcommand{\thefootnote}{\fnsymbol{footnote}}
\section{Introduction}
\label{intro}

Neutrinos are unique astronomical messengers. 
The observation of extra-galactic high energy astrophysical neutrinos
would imply a hadronic origin of cosmic rays.  Moreover, 
unlike photons or 
charged particles, they travel across the Universe without 
deflection by interstellar magnetic fields or absorption by
intervening matter. 
Existing and upcoming neutrino detectors (see for example~\cite{AMANDA,IceCube0,
BAIKAL,ANTARES,RICE,ANITA,KM3}) are expected to eventually observe high-energy neutrinos 
from Active Galactic Nuclei, Gamma Ray Bursts, GZK
processes and other feasible sources. 

High-energy cosmic neutrinos are also unique messengers of physics 
of and beyond the Standard Model. 
With a typical baseline of inter-galactic scales, neutrinos propagate incoherently
such that the transition probabilities between the flavor eigenstates are 
described only by the elements of the lepton mixing matrix.
The flavor composition at the Earth 
thus carries important information on the 
lepton flavor structure~\cite{FlavorR}, see \cite{sandip} for a
recent review. 
Furthermore, the long baselines and high energies allow for
interesting discussions of exotic possibilities such as 
neutrino decay~\cite{Decay}, pseudo-Dirac neutrinos \cite{pDirac}, 
Lorentz and CPT violation~\cite{LV}, which may show an effect in flavor 
ratio deformations of the diffuse spectrum~\cite{BCGW}. 

Since the ultra-high energy neutrinos span a wide range of
energies, they can be sensitive to the 
Glashow Resonance (GR)~\cite{GR, Berezinsky:1977sf, Berezinsky:1981bt}, which refers to 
the resonant formation of an intermediate $W^-$ in $\bar{\nu}_e e$ collision 
at the anti-neutrino energy $E_{\bar{\nu}} = 6.3\,{\rm PeV} \simeq 10^{6.8}$
GeV. This is a particularly interesting process~\cite{GR0,GR1,GR2,GR3}, unique in its
sensitivity to only anti-neutrinos. In particular, because the
relative $\bar{\nu}_e$ content of $pp$ and $ p\gamma $ collision final states is very 
different, the question of which of these two processes lie at the origin of high energy
neutrinos can, in principle, be tested well with GR events. Indeed, 
earlier works have focused mainly on the resonance detection via 
shower events and on how the GR can be used as a discriminator of the 
relative abundance of the $pp$ and $p\gamma$ sources. 

\begin{figure}[t]
\begin{center}
\scalebox{0.5}{
\includegraphics{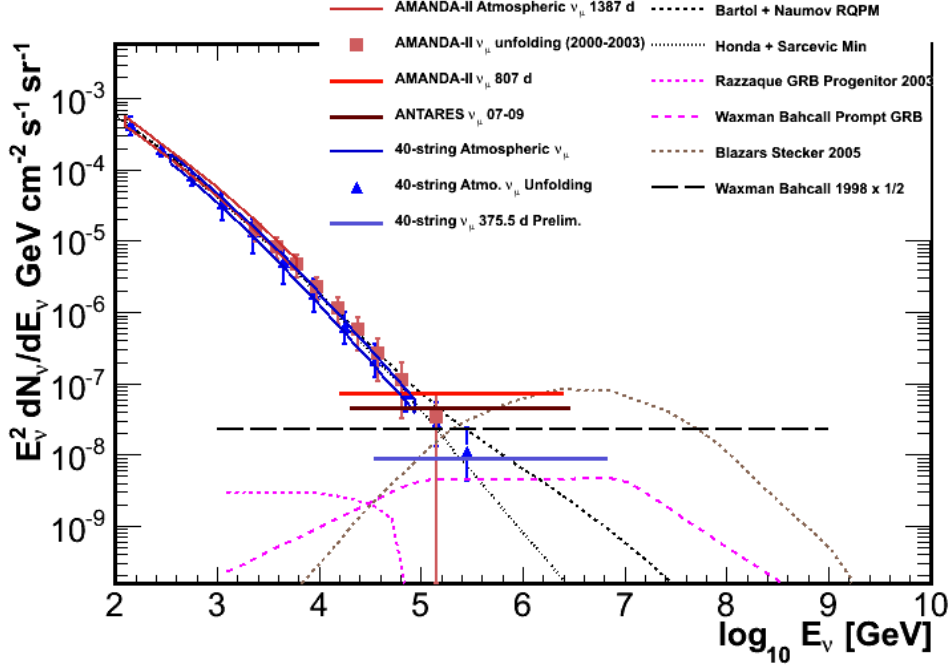}
}
\end{center}
\caption{Present experimental bounds from IceCube on the diffuse \numu flux assuming 
an $E^{-2}$ injection spectrum at source~\cite{IC40}. 
Predictions of neutrino fluxes from several theoretical models are also shown.}
\label{fig:icecube-bounds}
\end{figure}

Our emphasis in this work is not just on the detectability of the resonance 
itself, but also on its feasibility as a tool to detect
the first extra-galactic diffuse neutrino signals. We 
recalculate expected GR event numbers and their dependence on the
relative contribution of $pp$ and $p\gamma$ sources. Our work
updates and generalizes the results of Ref.~\cite{GR0}. 
To calculate the number of events, we use the Waxman--Bahcall $E^{-2}$ 
spectrum~\cite{WB} as a benchmark neutrino spectrum.  
Recently IceCube, the construction of which has been completed in December 2010 with 
86 strings, has improved the upper bound of the cosmic neutrino flux~\cite{IC40}. 
The current limits on the diffuse neutrino flux are shown 
in Fig.~\ref{fig:icecube-bounds}. 
If the neutrino flux is to be observed, it is reasonable to assume that 
it will emerge above the atmospheric background while staying below the current 
experimental upper bounds.
The present status of these limits leads us to believe that this is likely to happen 
at energies of $10^6$ GeV or greater, close to region of the Glashow resonance.
Therefore, it is useful and timely to revisit this resonance 
region carefully to reassess its potential as a tool to detect the cosmic diffuse 
neutrinos. 

\begin{figure}
 \begin{minipage}{0.5\hsize}
  \begin{center}
  \scalebox{0.8}{ \includegraphics[width=70mm]{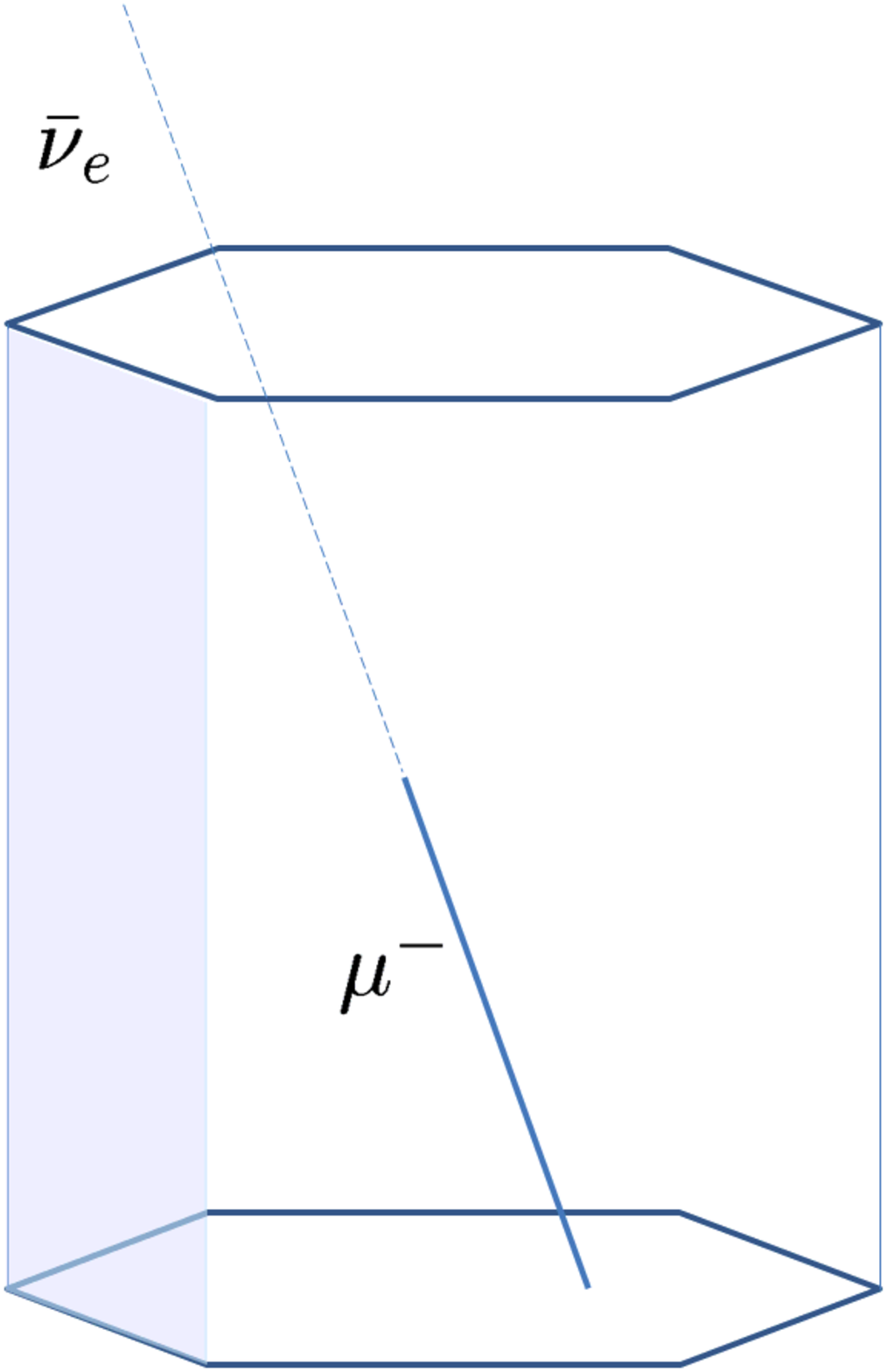}}
  \end{center}
  \caption{Pure muon}
  \label{fig:one}
 \end{minipage}
 \begin{minipage}{0.5\hsize}
  \begin{center}
  \scalebox{0.8}{\includegraphics[width=70mm]{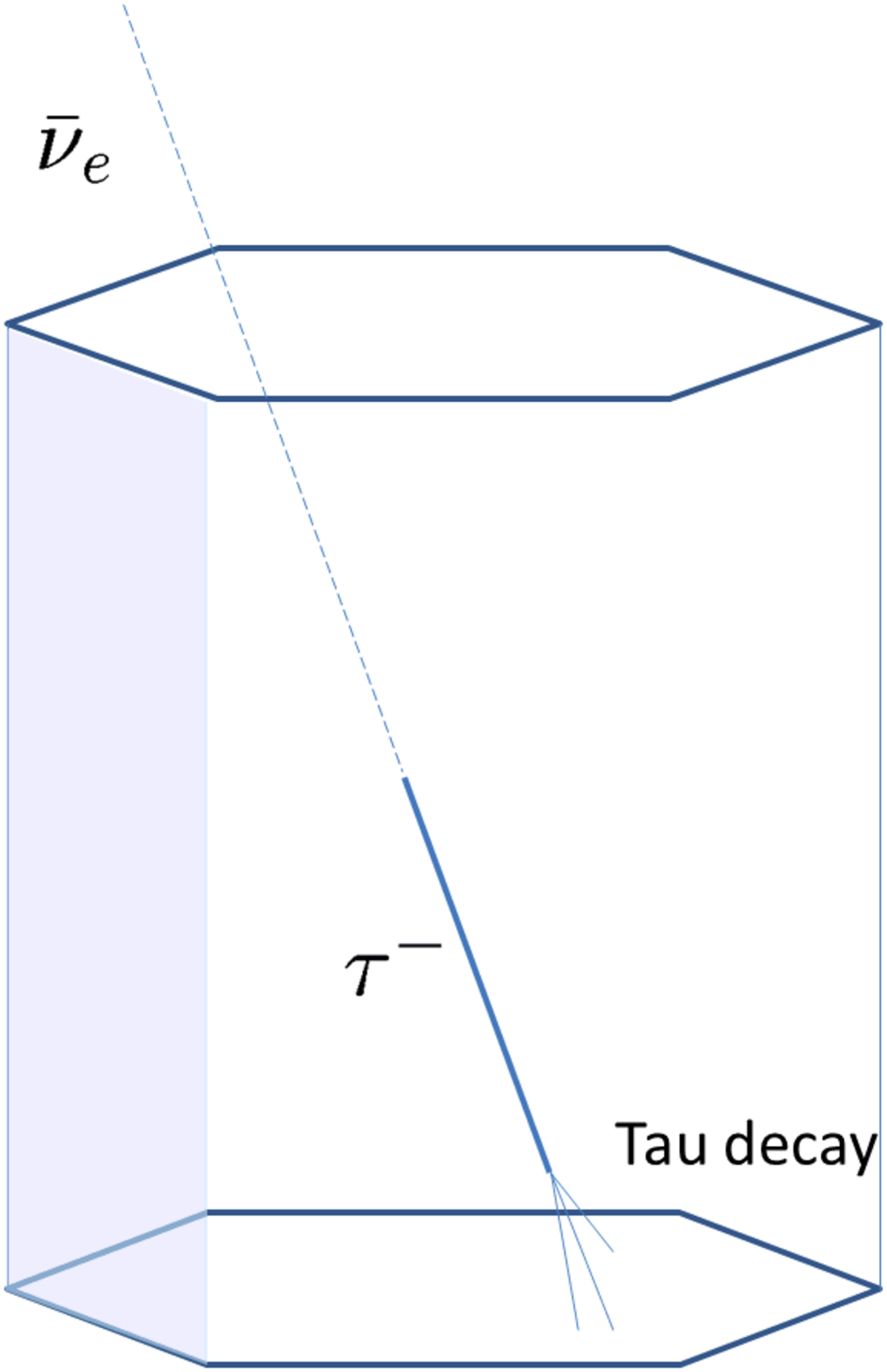}}
  \end{center}
  \caption{Contained lollipop}
  \label{fig:two}
 \end{minipage}
\end{figure}

In addition, we point out that there are two types of distinctive resonant processes 
besides the standard shower signatures from $\bar{\nu}_e e \to {\rm hadrons}$ and 
$\bar{\nu}_e e \to \bar{\nu}_e e$ considered in the literature. 
We call these new signatures ``pure muon''  and ``contained lollipop'' events. 
A pure muon event occurs when only a muon track (and nothing else) is created 
inside the detector volume by the resonant process $\bar{\nu}_e e \to
\bar{\nu}_\mu \mu$. We sketch the signature in Fig.~\ref{fig:one}. 
Unlike the neutrino--nucleon charged current scattering
$\nu_\mu N \to \mu X$ (and its charge conjugated counterpart), 
the pure muon track is not accompanied by any shower activity
at its starting point. 
We note that in $\nu_\mu N \to \mu X$ processes with PeV neutrino energies, 
about $26$\% of the initial neutrino energy is transfered to the
kicked quark, which turns into a hadronic cascade~\cite{xsec}. 
Thus, a muon track from $\nu_\mu N \to \mu X$ is accompanied by a $\sim 200\,{\rm m}$
radius shower at the interaction vertex for ${\rm PeV}$ neutrino
energies. This is clearly distinguishable from the muons of the pure
muon event $\bar{\nu}_e e \to \bar{\nu}_\mu \mu$. 
A possible background against this signal is the non-resonant electroweak 
process $\nu_\mu e \to \mu \nu_e$.
The cross section is however three orders of magnitude smaller than $\bar{\nu}_e e 
\to \bar{\nu}_\mu \mu$ at the resonant energy.
The pure muon is therefore essentially background free in the
neighborhood of the resonance energy and even one event
implies discovery of the resonance and signals the presence of diffuse extra-galactic 
flux.

A contained lollipop event occurs for $\bar{\nu}_e e \to
\bar{\nu}_\tau \tau$: a tau is created and decays 
inside the detector with a sufficient length of the tau track, see
Fig.~\ref{fig:two}. 
Again, due to the lack of shower activity at the initial vertex, the contained 
lollipop is also clearly separated from the standard double bang~\cite{DB}
signature induced by the $\nu_\tau N + \bar{\nu}_\tau N $ charged
current scattering, and it is therefore also essentially free from
background. \\

This paper is organized as follows: 
in Section~\ref{x-sec}, we briefly review the cross sections associated with the GR, 
in Section~\ref{flux}, we discuss the expected neutrino flux 
for $pp$ and $p\gamma$ sources, keeping their relative flux ratio as a free parameter. 
In Section~\ref{event}, the event rate and the signal-to-background ratio are studied, 
and finally, conclusions are presented in Section~\ref{conc}.

\section{The Glashow-resonance and its relevance to present day UHE neutrino 
detection}
\label{x-sec}
Ultra-high energy electron anti-neutrinos allow the resonant formation of $W^-$ in 
their interactions with electrons, at $6.3$ PeV. This process, known as the Glashow 
resonance~\cite{GR, Berezinsky:1977sf, Berezinsky:1981bt} has, in the resonance energy 
band, several notably high cross-sections for the allowed decay channels of the $W^-$. 
In particular, the differential cross-section for $\bar{\nu}_e e \to \bar{\nu}_\mu \mu$ 
is given by 
\begin{equation}
	\frac{d\sigma}{dy}
	\left( 
		\bar{\nu}_e e \to \bar{\nu}_\mu \mu
	\right)
	=
	\frac{G_F^2 m E_\nu}{2\pi} 
	\frac{4(1-y)^2
	\bigl( 
		1 - (\mu^2 - m^2)/2m E_\nu
	\bigr)^2}
	{\left(
		1 - 2 m E_\nu/M_W^2
	\right)^2 + \Gamma_W^2/M_W^2},
\end{equation}
and, for hadrons one may write
\begin{equation}
	\frac{d\sigma}{dy}
	\left( 
		\bar{\nu}_e e \to \text{hadrons}
	\right)
	=
	\frac{d\sigma}{dy}
	\left( 
		\bar{\nu}_e e \to \bar{\nu}_\mu \mu
	\right)
	\times
	\frac{\Gamma\left(W \to \text{hadrons}\right)}
		 {\Gamma\left(W \to \bar{\nu}_\mu \mu\right)}.
\end{equation}
The above expressions hold in the lab frame where $m = $ electron mass, $\mu = $ muon 
mass, $M_W = W^-$ mass, $y=E_\mu/E_\nu$, and $\Gamma_W$ is the total width of the $W$.

Table~\ref{tab1}~\cite{xsec} lists the total cross-sections at $E_\nu^{\text{res}} 
= 6.3$ PeV. We note that for the leptonic final states, one expects (very nearly) equal 
cross-sections regardless of whether one produces $\antinumu \mu$, $\antinutau \tau$ or 
$\antinue e$.

In the right panel of Table~\ref{tab1} we list, also at $E_\nu = 6.3$ PeV, 
the possible non-resonant 
interactions which could provide backgrounds to the interactions listed in 
the left panel of Table~\ref{tab1}. We note that the total resonant cross-section, 
$\antinue e \to 
\text{anything}$ is about $360$ times higher than the total neutrino-nucleon 
cross-section, $\numu N \to \mu + \text{anything}$. The cross-section for $\antinue e 
\to \text{hadrons}$ is about $240$ times its non-resonant hadron producing background 
interaction $\numu N \to \mu + \text{anything}$. Even the resonant leptonic final state 
interactions have cross-sections about 40 times that of the total $\numu N \to \mu + 
\text{anything}$ cross-section. Finally we note that the ``pure-muon" and ``contained 
lollipop" resonant processes discussed in the Sec.~\ref{intro} have negligible 
backgrounds. For example, the process $\antinue e \to \antinumu \mu$ (pure muon) has 
a cross-section about $1000$ times higher than its non-resonant counterpart $\numu e \to 
\nue \mu$.

Given these considerations and the fact that the present bounds shown 
in Fig.~\ref{fig:icecube-bounds} restrict observational diffuse fluxes to energies 
above $10^6$ GeV (\textit{i.e.}, close to the GR region), the GR, in spite of its 
narrow span of energy, may be an important discovery tool for the yet
to be observed extra-galactic diffuse neutrino spectrum.

\begin{table}
 \begin{minipage}{0.5\hsize}
  \begin{center}
  \begin{tabular}{ccc}\hline
Interaction & & $\sigma \,[\rm cm^2]$ \\ 
\hline 
$\antinue e \to \antinue e$  & & $5.38 \times 10^{-32}$ \\ 
 
$\antinue e \to \antinumu \mu$  & & $5.38 \times 10^{-32}$\\ 
 
$\antinue e \to \antinutau \tau$  & & $5.38 \times 10^{-32}$ \\ 
 
$\antinue e \to \text{hadrons}$  & & $3.41 \times 10^{-31}$\\ 
 
$\antinue e \to \text{anything}$  & & $5.02 \times 10^{-31} $ \\ 
\hline 
\end{tabular}
  \end{center}
 \end{minipage}
 \begin{minipage}{0.5\hsize}
  \begin{center}
  \begin{tabular}{ccc}\hline
Interaction & & $\sigma\, [\rm cm^2]$ \\ 
\hline 
$\numu N \to \mu + \text{anything}$ & & $1.43 \times 10^{-33}$ \\ 
 
$\numu N \to \numu + \text{anything}$ & & $6.04 \times 10^{-34}$ \\ 
 
$\numu e \to \nue \mu $  & & $5.42 \times 10^{-35}$ \\ 
\hline 
\end{tabular}
\vspace{10mm}
  \end{center}
 \end{minipage}
\caption{Resonant GR cross-sections for electron anti-neutrino~(left panel) 
and non-resonant~(right panel) interactions at $E = 6.3$ PeV.}
\label{tab1}
\end{table}

\section{Diffuse Neutrino Fluxes for $pp$ and $p\gamma$ sources}
\label{flux}
The search for cosmic neutrinos with PeV energies is motivated by 
observations of cosmic rays.
It has been conjectured that cosmic ray engines accelerate protons
and confine them with magnetic fields in the acceleration region.
The accelerated protons interact with ambient photons or protons, 
producing neutrons and charged pions. Charged particles are trapped
by magnetic fields, while neutral particles escape from the
source region, decay and produce observable cosmic rays and neutrinos.
If the source region is optically thin, the energy density of neutrinos
scales linearly with the cosmic ray density and the neutrino intensities
are co-related with the observed cosmic ray flux. 

The result of these considerations for the expected total neutrino 
flux (the sum over all species) at the source is the Waxman-Bahcall
flux, given by~\cite{WB} 
\begin{eqnarray}
E_\nu^2 \Phi_{\nu + \bar{\nu}} \,=\,
2\times 10^{-8} \epsilon_\pi \xi_z \quad (\rm GeV\, cm^{-2}\, s^{-1}\,
sr^{-1} \,). 
\label{WB}
\end{eqnarray}
Here $\xi_z$ is a function of the red-shift parameter $z$ alone,
representing the evolution of sources with red-shift, and $\epsilon_\pi$ is the
ratio of pion energy to the emerging nucleon energy at the source. One
has $\xi_z \approx 0.6$ for no source evolution, while $\xi_z \approx 3$ for an evolution 
$\propto (1+z)^3$.
Depending on the relative ambient gas and photon densities,
the neutrino production originates in either $p\gamma$ or $pp$
interactions. For the $pp$ case $\epsilon_\pi \approx 0.6$ and for the
$p\gamma$ case $\epsilon_\pi \approx 0.25$. 

Since source distributions and types are not well known, we parameterize the
relative $pp$ and $p\gamma$ contributions to the total flux with a dimensionless 
parameter $x$, $0\leq x \leq 1$, so that
\begin{eqnarray} \label{main}
\Phi_{\rm source} = x\Phi^{pp}_{\rm source} + (1-x)\Phi^{p\gamma}_{\rm source},
\end{eqnarray}
where $\Phi^{pp/p\gamma}$ represents the neutrino flux from $pp/p\gamma$ interactions.
We assume here that neutron decays, which (as discussed in \cite{GR2}) could be 
present in certain sources give negligible contributions to the overall flux. 
Effects like multi-pion processes producing $\pi^-$ events in $p\gamma$ sources, 
can be included in the parameterization.

The flavor composition at the source is given by
$(\nu_e,\nu_\mu,\nu_\tau) = (\bar{\nu}_e,\bar{\nu}_\mu,\bar{\nu}_\tau) \approx 
(1,2,0)$ for a $pp$ source and
$(\nu_e,\nu_\mu,\nu_\tau) \approx (1,1,0)$ and
$(\bar{\nu}_e,\bar{\nu}_\mu,\bar{\nu}_\tau) \approx (0,1,0)$
for the $p\gamma$ case. 
These configurations are changed by the incoherent propagation from the
source to earth. 
The transition probabilities between flavor eigenstates are described by 
three mixing angles and one CP violating phase.
By using $\theta_{12} = 35^\circ$, $\theta_{13}= 0$, and $\theta_{23}= 45^\circ$
as reference values of the lepton mixing angles, the flavor ratios at the earth
become $(\nu_e,\nu_\mu,\nu_\tau) = (\bar{\nu}_e,\bar{\nu}_\mu,\bar{\nu}_\tau) = 
(1,1,1)$ for $pp$, while 
$(\nu_e,\nu_\mu,\nu_\tau) = (0.78,0.61,0.61)$ and
$(\bar{\nu}_e,\bar{\nu}_\mu,\bar{\nu}_\tau) = (0.22,0.39,0.39)$ for 
fluxes from $p\gamma$ interactions.
Finally, the flux for each neutrino species is given by
\begin{eqnarray}
&&E_\nu ^2 \Phi_{\nu_e} = 2\times10^{-8} \xi_z \left[ \,
x \,\frac{1}{6}\cdot 0.6 + (1-x)\,\frac{0.78}{3}\cdot 0.25 \right],
\label{f1}\\
&&E_\nu ^2 \Phi_{\nu_\mu} = 2\times10^{-8} \xi_z \left[ \,
x \,\frac{1}{6}\cdot 0.6 + (1-x)\,\frac{0.61}{3}\cdot 0.25 \right] 
= E_\nu ^2 \Phi_{\nu_\tau},
\label{f2}\\
&&E_\nu ^2 \Phi_{\bar{\nu}_e} = 2\times10^{-8} \xi_z \left[ \,
x \,\frac{1}{6}\cdot 0.6 + (1-x)\,\frac{0.22}{3}\cdot 0.25 \right],
\label{f3}\\
&&E_\nu ^2 \Phi_{\bar{\nu}_\mu} = 2\times10^{-8} \xi_z \left[ \,
x \,\frac{1}{6}\cdot 0.6 + (1-x)\,\frac{0.39}{3}\cdot 0.25 \right] 
= E_\nu ^2 \Phi_{\bar{\nu}_\tau},
\label{f4}
\end{eqnarray}
in units of ${\rm GeV\, cm^{-2}\, s^{-1}\, sr^{-1} \,}$.
Note that in our phenomenological analysis we are starting with Eq.~(\ref{main}),
{\it i.e.,} we do not specify the mechanism generating the original protons
and/or photons.
Therefore, we can use a common energy in the above relations~(\ref{f1})-(\ref{f4}).
The equalities between $\nu_\mu$ and $\nu_\tau$ flavors, both for neutrinos and 
anti-neutrinos, are the consequence of  vanishing $\theta_{13}$
(actually, vanishing of the real part of $U_{e3}$ would suffice) 
and maximal $\theta_{23}$ used in the calculation.
The uncertainty in $\theta_{13}$ and $\theta_{23}$ breaks this equality and 
changes the fluxes by about $10$ to $20$ \% for the $2\sigma$ allowed ranges
of the mixing angles~\cite{3gen}. 
Note that the total intensity becomes maximal for the pure $pp$ case $x=1$.
With a strong evolution value $\xi_z = 3$, the maximal value is $\sum_\alpha E_\nu^2 
\Phi_{\nu_\alpha + \bar{\nu}_\alpha} = 
3.6 \times 10^{-8} \,{\rm GeV\, cm^{-2}\, 
s^{-1}\, sr^{-1} \,}$, which agrees with the latest upper bound on the $E^{-2}$ 
spectrum~\cite{IC40}.
In what follows, we use these fluxes with $\xi_z = 3$ as an example to
calculate the event rates.

\section{Event rates and signal/background ratio}
\label{event}
As discussed before, we will look at both shower and muon/tau-track
events to identify 
unique signatures for cosmic neutrinos via the Glashow resonance. 
In this context, we first focus on the shower events. 

\subsection{Shower signatures of the Glashow resonance}

Among the resonance processes, it turns out that the only 
channel significantly contributing to the events is the hadronic interaction 
$\bar{\nu}_e e \to {\rm hadrons}$, while the contributions from 
the other channels are negligibly small.
Beside the hadronic channel, the following two decay modes produce
electromagnetic showers in the detector;
~$i)$ $\bar{\nu}_e e \to \bar{\nu}_e e$ and~$ii)$ 
$\bar{\nu}_e e \to \bar{\nu}_\tau \tau$ with
$E_\tau \lesssim 2\, {\rm PeV}$.
A tau of $E_\tau \gtrsim 2\,{\rm PeV}$ travels more than
$100\,{\rm m}$ before decay and can be separated from a 
single shower\footnote{This is identified as the contained lollipop if the
shower provided by the tau decay occurs inside the detector volume.}.
Notice that the hadronic channel constitutes $68$\% of the total decay width of $W^-$, 
whereas $i)$ and $ii)$ constitute $11$\% each.
Furthermore, only half of the parent neutrino energy becomes shower
energy in $i)$ and $ii)$,
while all energy is converted to shower energy in the hadronic mode.

The event rate of $\bar{\nu}_e e \to {\rm hadrons}$ 
is calculated as 
\begin{eqnarray}
{\rm Rate} \,=\,
2\pi \,\frac{10}{18}N_A V_{\rm eff} \!\int \!\! dE_\nu \int_0^1\!\! dy \,
\frac{d\sigma}{dy}(\bar{\nu}_e e \to \text{hadrons})\,\Phi_{\bar{\nu}_e}(E_\nu),
\end{eqnarray}
where $N_A = 6.022 \times 10^{23} \,{\rm cm^{-3}}$ and $V_{\rm eff} 
\approx 2\,{\rm km^3}$.
The effective volume is taken as twice as large as the instrumental volume
since the radius of the showers with the resonant energy is about $300\,{\rm m}$.
The events are integrated over the upper half sphere since up-moving electron neutrinos
are attenuated by the earth matter.
At the resonance peak, the integrated cross section is $3.4\times 10^{-31}\,{\rm cm^2}$.
With the $pp$ ($p\gamma$) source flux $E_\nu^2 \Phi_{\bar{\nu}_e} 
= 6\,(1.1) \times 10^{-9}\,{\rm GeV\, 
cm^{-2}\, s^{-1}\, sr^{-1} \,}$, $3.2$ ($0.6$) events are expected at the
resonant energy region for 1 year of observation.
\begin{figure}[t]
	\centering
	\includegraphics[scale=0.205]{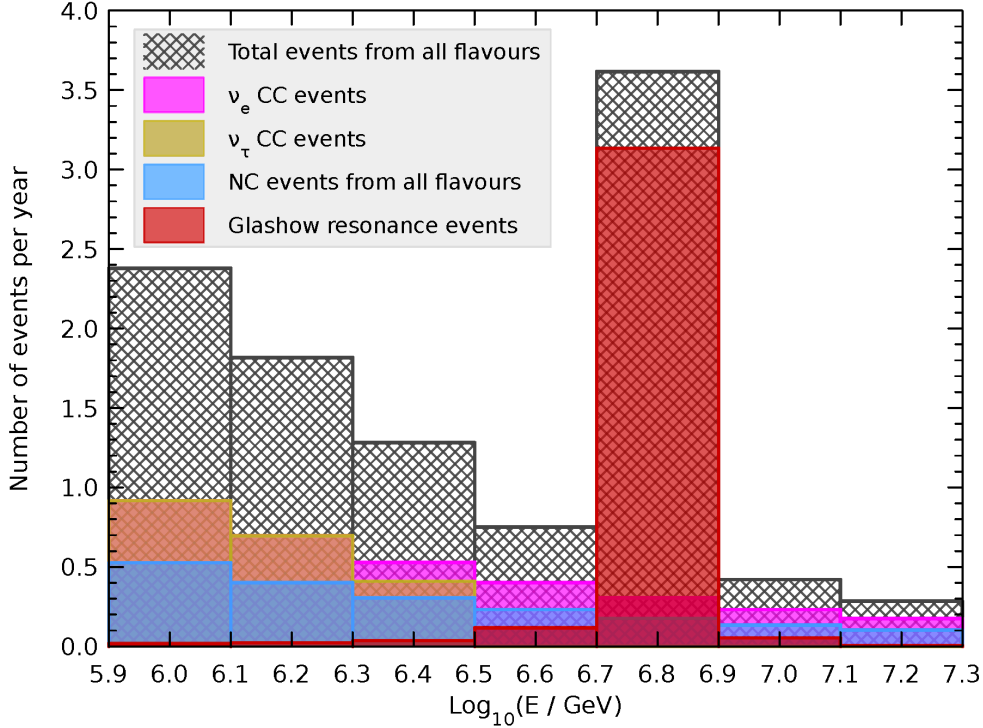}
	\caption{\label{A}The shower spectrum for pure $pp$ sources, $x=1$. We have neglected events from the interactions $\antinue e \to \antinue e$ and $\antinue e \to \antinutau \tau$ which contribute, comparatively, a very tiny fraction of events to the spectrum.}
\end{figure}
The off-resonant background events receive contributions from 
$\nu_e N + \bar{\nu}_e N$ (CC)
and $\nu_\alpha N + \bar{\nu}_\alpha N$ (NC), where CC (NC) represents
the charged (neutral) current.
The tau contribution $\nu_\tau N + \bar{\nu}_\tau N$ (CC) is irrelevant at the 
resonance energy bin since a tau with $E_\tau \gtrsim 2\,{\rm PeV}$ manifests 
itself as a track.  The event rate of $\nu_e N + \bar{\nu}_e N$ (CC) is given by
\begin{eqnarray}
{\rm Rate} =
 2\pi\, N_A V_{\rm eff} \!\int\!\! dE_\nu \left[\,
 \sigma_{\rm CC}(\nu N) \,\Phi_{\nu_e}(E_\nu)
+ \sigma_{\rm CC}(\bar{\nu} N) \,\Phi_{\bar{\nu}_e}(E_\nu)\,\right],
\end{eqnarray}
where $ \sigma_{\rm CC}(\nu N/ \bar{\nu} N)$ is the neutrino--nucleon cross
section which is $\approx 1.4\times 10^{-33}\,{\rm cm^2}$ at $E_\nu = 
6.3\,{\rm PeV}$ \cite{xsec}.   
For $\nu_\alpha N + \bar{\nu}_\alpha N$ (NC), 
the rate is calculated as 
\begin{eqnarray}
{\rm Rate} \,\simeq\,
2\pi\, N_A V_{\rm eff} \sum_{\alpha=e,\mu,\tau}\,
\!\int_{E_0/\langle y \rangle}^{E_1/\langle y \rangle} \!\! dE_\nu \left[\,
 \sigma_{\rm NC}(\nu N) \,\Phi_{\nu_\alpha}(E_\nu)
+ \sigma_{\rm NC}(\bar{\nu} N) \,\Phi_{\bar{\nu}_\alpha}(E_\nu)\, \right],
\label{NC}
\end{eqnarray}
for the shower energy between $E_0$ and $E_1$. Here $\langle y \rangle$ is
the mean inelasticity which is well described by the average value 
$\langle y \rangle = 0.26$ at PeV energies.
The NC cross section at the resonant peak is $\approx 6 \times 10^{-34}\,{\rm cm^2}$.
In the NC process, only a part of the neutrino energy (about $26$\%) is
converted to shower energy, 
so that the NC contribution is generally small with respect to the CC
event number. 
We have assumed $100$\% shadowing by the earth for the sake of simplicity,
but note that muon and tau neutrinos are not completely attenuated and 
actually about $20$\% of them survive in average at the resonant energy.
The muon and tau component in Eq.~(\ref{NC}) would thus receive 
$\simeq 20$\% enhancement in a more precise treatment.
For showers with energies $10^{6.7}\,{\rm GeV}< E_{\rm shower} < 10^{6.9}\,{\rm GeV}$,
for example, the rate reads $0.31\,{\rm yr^{-1}}$ for CC and
$0.18\,{\rm yr^{-1}}$ for NC in the case of a $pp$ flux.
\begin{figure}[t]
	\centering
	\includegraphics[scale=0.205]{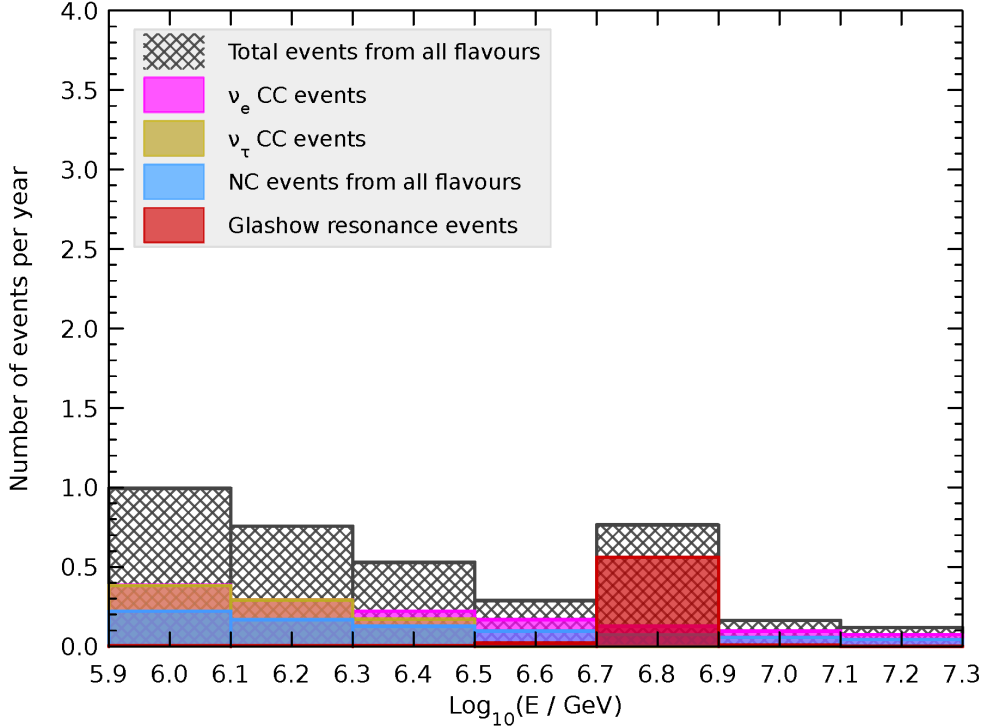}
	\caption{\label{B}The shower spectrum for pure $p\gamma$ sources, $x=0$. We have neglected events from the interactions $\antinue e \to \antinue e$ and $\antinue e \to \antinutau \tau$ which contribute, comparatively, a very tiny fraction of events to the spectrum.}
\end{figure}

\begin{figure}
\begin{center}
\scalebox{0.72}{
\includegraphics{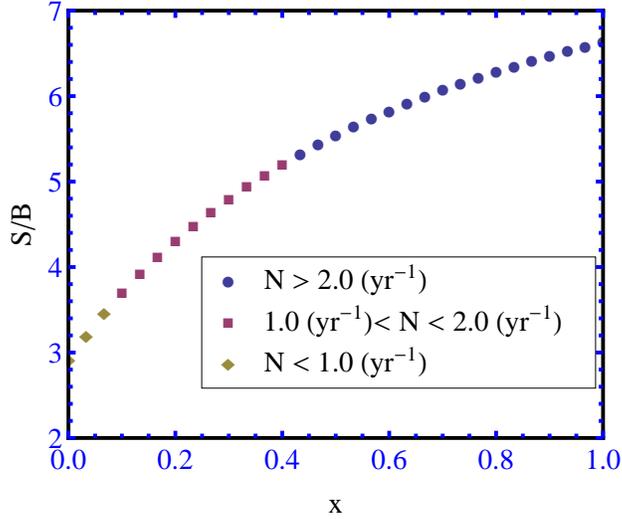}
}
\end{center}
\caption{The ratio of $\bar{\nu}_e e \to {\rm hadrons}$
to the off-resonant processes in the resonant bin 
as a function of $x$. $N$ represents the total number of
event in the resonant bin.}
\label{sb}
\end{figure}

Figs.~\ref{A} and~\ref{B} show the number of events in the neighborhood of
the resonant energy.
Fig.~\ref{A} is for a pure $pp$ flux with $x=1$ and Fig.~\ref{B} 
for a pure $p\gamma$ flux $x=0$.
As was pointed out in~\cite{GR0}, the resonance peak is clearly seen
for a pure $pp$ source, whereas the peak is significantly weakened
for $p\gamma$ sources. 
We have divided the energy decade $10^{6.3}\,{\rm GeV} < E_{\rm shower} < 
10^{7.3}\,{\rm GeV}$ into five bins by assuming the energy resolution of
the shower to be $\log_{10}(E_{\rm shower}/{\rm GeV}) = 0.2$. 
Notice that $\nu_\tau N + \bar{\nu}_\tau N$ and $\nu_e N + \bar{\nu}_e
N$ generate the same event numbers at low energies in Fig.~\ref{A},
since the cross section and the $pp$ fluxes are flavor blind. 
For energies higher than $10^{6.5}\,{\rm GeV}$, events numbers from
$\nu_\tau N + \bar{\nu}_\tau N$ are lower because the tau track becomes
visible and the events can be separated from a single shower. 
Fig.~\ref{sb} shows the ratio of $\bar{\nu}_e e \to {\rm hadrons}$
to the sum of all off-resonant processes in the resonant bin
$10^{6.7}\,{\rm GeV}< E_{\rm shower} < 10^{6.9}\,{\rm GeV}$ as a function
of $x$. The ratio rises from $3$ at $x=0$ to about $7$ at $x=1$.

While the total spectral shape shown in Fig.~\ref{A} and~\ref{B}
crucially depends on the parameter $x$, it also depends on the flavor 
composition at the earth. 
For example, if the muon and tau components would evanesce while the 
(anti-)electron would stay constant, perhaps due to non-standard
physical effects affecting the oscillation probabilities, 
the ratio of the resonant to off-resonant events 
is enhanced over the ``standard'' maximal value set by $x=1$. 
In an opposite case where only the electron component is damped, the ratio
would be anomalously small. 
Hence the shower spectral shape around the resonance has certain sensitivities
to the deformation of the flavor composition, being a complementary test 
to the shower/muon track ratio. This issue is separately studied in~\cite{fu}.

\subsection{Novel signatures of the Glashow resonance} 

We now discuss other unique signatures of the Glashow
resonance; the pure muon and the contained lollipop.
If the resonant process $\bar{\nu}_e e \to \bar{\nu}_\mu \mu$ takes place in the
detector volume, it will be observed as a muon track without shower activities
at its starting point, see Fig.~\ref{fig:one}. 
This ``pure muon'' signature will be clearly distinguishable from the usual muon 
track from $\nu_\mu N$ charged current interactions. 
The probability that the shower associated with the $\nu_\mu N$ CC
process does not reach the detection threshold is extremely  
small at PeV energies. There is a possibility that bremsstrahlung of the pure muon 
may distort the signal. However, this bremsstrahlung occurs only about $10$\% 
of the time, and the energy fraction carried by the radiation is much 
smaller than $\langle y \rangle = 0.26$ of the shower. 
Therefore the probability that the signal is misidentified as the $\nu_\mu N \to \mu X$
is expected to be small. 
The only remaining candidate for background is thus the muon created by the non-resonant 
process $\nu_\mu e \to \mu \nu_e$.

\begin{figure}[t]
\begin{center}
\scalebox{0.62}{
\includegraphics{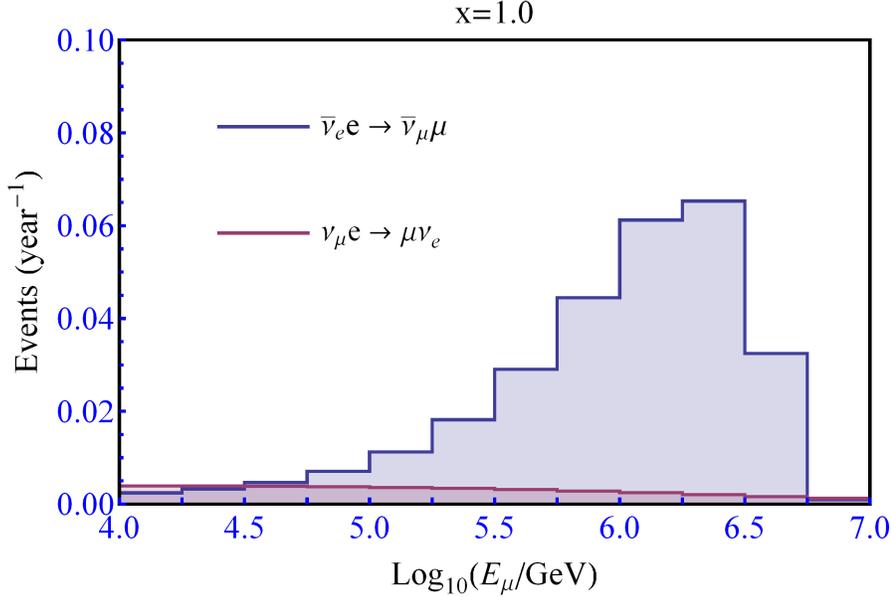}
}
\end{center}
\caption{The number of pure $\mu$ events as the functions of
the muon energy for a pure $pp$ source, $x=1$.}
\label{C}
\end{figure}

The event rate of $\bar{\nu}_e e \to \bar{\nu}_\mu \mu$ with the muon
energy $E_0 < E_\mu < E_1$ is calculated by
\begin{eqnarray}
{\rm Rate} =
2\pi\,\frac{10}{18}N_A V \!\left[
\int_{E_0}^{E_1} \!\! dE_\nu \int_{\frac{E_0}{E_\nu}}^1\!\! dy \,
+ 
\int_{E_1}^{\infty} \!\! dE_\nu \int_{\frac{E_0}{E_\nu}}^{\frac{E_1}{E_\nu}}\!\! dy \,
\right]
\frac{d\sigma}{dy}(\bar{\nu}_e e \to\bar{\nu}_\mu \mu )\,\Phi_{\bar{\nu}_e}(E_\nu),
\end{eqnarray}
where $V = 1\,{\rm km^3}$ is the instrumental volume of IceCube.
The non-resonant process $\nu_\mu e \to \mu \nu_e$ is also calculated in
the same manner by replacing the cross section and the flux.

Fig.~\ref{C} shows the event number spectrum of these processes.
It is seen that $\bar{\nu}_e e \to \bar{\nu}_\mu \mu$ is dominant
in the energy regime $5.0 < \log_{10}(E_\mu/{\rm GeV}) < 6.75$, where
the $\nu_\mu e \to \mu \nu_e$ contribution is tiny for $x=1$. 
The integrated number of resonant events in this region is $0.26\,{\rm
yr^{-1}}$. Although the absolute number of the expected event is small,
even a single detection of the pure muon event becomes essentially a discovery
of the resonance at this energy regime due to its uniqueness. For $x=0$, 
the rate decreases to $0.048\,{\rm yr^{-1}}$.

\begin{figure}
\begin{center}
\scalebox{0.62}{
\includegraphics{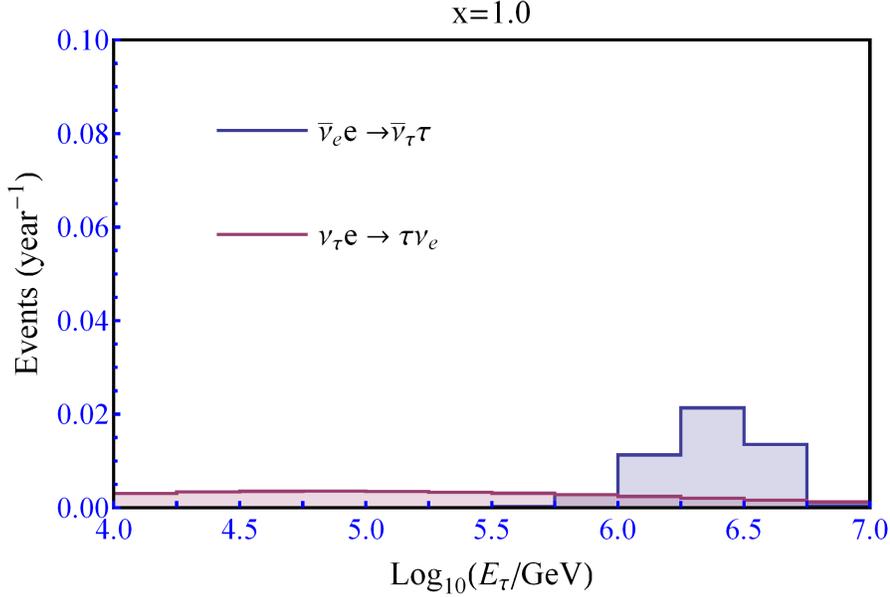}
}
\end{center}
\caption{The event spectrum of the contained lollipop for a pure $pp$
source, $x=1$.}
\label{D}
\end{figure}

Turning to the contained lollipop, this signature denotes the case
when the resonant process $\bar{\nu}_e e \to \bar{\nu}_\tau \tau$ takes place in the
detector volume and the tau decays a significant distance
thereafter, see Fig.~\ref{fig:two}. This will be observed as a tau track popping up 
inside the detector (without an initial hadronic shower) and a subsequent 
shower when it decays at the end of the track. 
It is a ``double-bang without the first bang'' so to speak.
The event rate with the tau energy of $E_0 < E_\tau < E_1$ is given by
\begin{eqnarray}
{\rm Rate} &=&
2\pi\,\frac{10}{18}N_A A \!\left[
\int_{E_0}^{E_1} \!\! dE_\nu \int_{\frac{E_0}{E_\nu}}^1\!\! dy \,
+ 
\int_{E_1}^{\infty} \!\! dE_\nu \int_{\frac{E_0}{E_\nu}}^{\frac{E_1}{E_\nu}}\!\! dy \,
\right]
\frac{d\sigma}{dy}(\bar{\nu}_e e \to\bar{\nu}_\tau \tau )\,\Phi_{\bar{\nu}_e}(E_\nu)
\nonumber\\
&&
\quad\quad\quad\quad\quad\quad\quad\quad\quad
\quad\quad\quad
\times \int_{L_0}^{L_1 - x_{\rm min}} \!\!\!\! dx_0 \int_{x_0 + x_{\rm min}}^{L_1} 
\!\!\!\! dx
\,\frac{1}{R_\tau}e^{-\frac{x - x_0}{R_\tau}}, 
\end{eqnarray}
where $R_\tau$ is the tau range
$R_\tau \,\simeq\,c\tau yE_\nu/m_\tau$,
and $A \approx 1 \,{\rm km^2}$ is the effective area of the detector, 
$L_1 - L_0 = L= 1\,{\rm km}$ is the length of the detector, 
$x_0$ is the neutrino interaction point, and
$x_{\rm min}$ is the minimum length to separate the tau decay point from 
the tau creation point. 
We take $x_{\rm min} = 100\,{\rm m}$ as a reference value.
The exponential factor accounts for the probability with which a tau
created at the point $x_0$ decays at the point $x$.

Fig.~\ref{D} shows the event spectrum for $x=1$ in comparison with
the obvious candidate of the background, $\nu_\tau e \to \tau \nu_e$.
The contained lollipop dominates in the $6.0 < \log_{10}(E_\tau/{\rm GeV}) < 6.75$ regime.
The integrated number of events in this region is $0.046\,{\rm yr^{-1}}.$
As the pure muon case, observation of a single event would essentially
become discovery of the resonance. Note however that the expected event
number is about five times smaller than the one from the pure muon
signature. 

\vspace{5mm}

Finally let us define the total signal of the Glashow resonance as the sum of
shower, muon track and contained lollipop events.
\begin{figure}
\begin{center}
\scalebox{0.62}{
\includegraphics{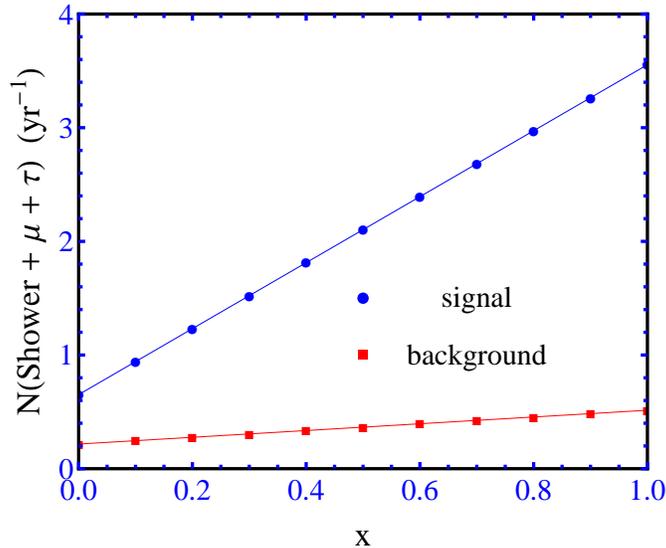}
}
\end{center}
\caption{Total number of the Glashow resonance signal as a function of $x$.
The lower (red) curve shows the background (\textit{i.e.},~off-resonant processes).}
\label{E}
\end{figure}
That is,
\begin{eqnarray}
N({\rm Shower} + \mu + \tau) \equiv N(\bar{\nu}_e e \to {\rm hadrons})
+ N(\bar{\nu}_e e \to \bar{\nu}_\mu \mu)
+ N(\bar{\nu}_e e \to \bar{\nu}_\tau \tau),
\end{eqnarray} 
where $N(\bar{\nu}_e e \to {\rm hadrons})$ is the number of shower events
in $6.7 < \log_{10}(E_{\rm shower}) < 6.9$ induced by $\bar{\nu}_e e
\to {\rm hadrons}$, $N(\bar{\nu}_e e \to \bar{\nu}_\mu \mu)$ is the
number of pure muon events in $5.0 < \log_{10}(E_\mu/{\rm GeV}) <
6.75$, and $N(\bar{\nu}_e e \to \bar{\nu}_\tau \tau)$ is the number of
contained lollipop events in $6.0 < \log_{10}(E_\tau/{\rm GeV}) < 6.75$.
Fig.~\ref{E} presents the total number of the GR events as a function
of $x$. 
The background (\textit{i.e.},~the off-resonant contributions) is defined by 
the summation of the total shower events other than $\bar{\nu}_e e \to {\rm hadrons}$
in $6.7 < \log_{10}(E_{\rm shower}) < 6.9$, 
the number of events for $\nu_\mu e \to \mu \nu_e$ in 
$5.0 < \log_{10}(E_\mu/{\rm GeV}) < 6.75$ and for 
$\nu_\tau e \to \tau \nu_e$ in $6.0 < \log_{10}(E_\tau/{\rm GeV}) < 6.75$.
The signal/background ratio rises from $\simeq 3$ at $x=0$ to $\simeq 7$ at $x=1$.
For $x=1$,  $7.2$ signal events against about $1$ background event are expected
with 2 years of data accumulation, which is well above the $99$\% C.L.~upper limit 
for the background only (observation of $1$ expected background event corresponds to
an upper limit of $5.79$ events at $99$\% C.L.~\cite{Stat}). 
For $x=0.5$, $6.3$ signal events and about $1$ background event is expected 
with 3 years of data accumulation. 
For the pure $p\gamma$ case $x=0$, $6.5$ signal and about $2$
background events are expected within 10 years of data accumulation, 
which is slightly below the $99$\% C.L.~upper limit for background
only observation (observation of $2$ expected background events
corresponds to an upper limit of $6.69$ events
at $99$\% C.L.~\cite{Stat}). Table~\ref{t1} shows the non-resonant, Glashow
resonance and total number of events for three representative values of $x$.
Depending on the relative abundance of the $pp$ and $p\gamma$ sources,
$20$, $12$ and $4$ events are expected in IceCube in $5$ years.  

Our focus in this section was on signatures and event numbers of the Glashow
resonance. From the more general point of view of discovery of high-energy cosmic 
neutrinos however, the off-resonant events (treated as backgrounds so far)
are also signals, being distinctive of neutrinos at energies which could not 
possibly be produced at any other neutrino source. 
Atmospheric neutrinos are not a significant background for such a discovery since 
their fluxes are negligibly low at PeV energies and their contribution, consequently, 
is insignificant. 

\begin{table}[t]
\begin{center}
\begin{tabular}{cccc}\hline\hline
$x$ &  Non-resonance & GR & Total \\\hline
0.0& 0.21 & 0.65 & 0.86 \\
0.5& 0.37 & 2.1 & 2.5 \\
1.0& 0.51 & 3.6 & 4.1 \\\hline
\end{tabular}
\caption{A list of expected numbers of events for 1 year data taking in IceCube.}
\label{t1}
\end{center}
\end{table}

\section{Conclusion}
\label{conc}
We have studied the Glashow resonance in the high-energy astrophysical neutrino
observatory IceCube. 
Besides the standard hadronic/electromagnetic cascade,
the pure muon from $\bar{\nu}_e e \to \bar{\nu}_\mu \mu$ and 
the contained lollipop signatures from $\bar{\nu}_e e \to
\bar{\nu}_\tau \tau$ were identified as clear signals of the
resonance. Applying a Waxman-Bahcall $E^{-2}$ flux in agreement with
recent limits, the event numbers for general $pp$ and $p\gamma$
sources were evaluated. 
If the neutrino fluxes are positioned with such intensities as
presently conjectured, the confirmation of the resonance
is possible with several years of data collection at IceCube.
The resonance could be used as a discovery tool for 
diffuse astrophysical neutrinos at PeV energies, and to obtain important information 
about cosmic-rays and astrophysical sources.

\subsection*{Acknowledgments}
We thank Walter Winter for helpful discussions. A.B.~would like to thank the Particle 
and Astroparticle Division of the Max-Planck-Institut f\"ur Kernphysik at Heidelberg 
for hospitality while the work was in progress. R.G.~would like to thank the University 
of Wisconsin phenomenology group and the CERN Theory Division for hospitality while 
this work was in progress. He would also like to acknowledge financial support from 
the DAE XI Plan Neutrino project.
W.R.~is supported by the ERC under the Starting Grant MANITOP and by the DFG in the 
project RO 2516/4-1 as well as in the Transregio 27. 
The work of A.W.~is supported by the Young Researcher Overseas Visits Program for 
Vitalizing Brain Circulation Japanese in JSPS.  
\bigskip


\end{document}